\documentclass[12pt]{article}

\usepackage{amsfonts,amssymb,bm,cite,amsmath}
\usepackage[dvips]{graphicx}

\newcommand {\beq}{\begin{equation}}
\newcommand {\eeq}{\end{equation}}
\newcommand {\beqa}{\begin{eqnarray}}
\newcommand {\eeqa}{\end{eqnarray}}

\begin{document}
\setlength{\oddsidemargin}{0cm}
\setlength{\baselineskip}{7mm}

\begin{titlepage}
\renewcommand{\thefootnote}{\fnsymbol{footnote}}

~~\\

\vspace*{0cm}
    \begin{Large}
       \begin{center}
         {A generalized volume law for entanglement entropy \\ on the fuzzy sphere}
       \end{center}
    \end{Large}
\vspace{1cm}

\begin{center}
           Mariko S{\sc uzuki}\footnote
            {
e-mail address : 
f5344003@gmail.com}
           {\sc and}
           Asato T{\sc suchiya}\footnote
           {
e-mail address : tsuchiya.asato@shizuoka.ac.jp}\\
      \vspace{1cm}

  {\it Department of Physics, Shizuoka University}\\
                {\it 836 Ohya, Suruga-ku, Shizuoka 422-8529, Japan}\\
         \vspace{0.3cm}     
         {\it Graduate School of Science and Technology, Shizuoka University}\\
               {\it 3-5-1 Johoku, Naka-ku, Hamamatsu 432-8011, Japan}

\end{center}

\hspace{5cm}

\begin{abstract}
\noindent
We investigate entanglement entropy in a scalar field theory on the fuzzy sphere.
The theory is realized by a matrix model. 
In our previous study, we confirmed
that entanglement entropy in the free case is proportional to
the square of the boundary area of a focused region. 
Here we argue that 
this behavior of entanglement entropy
can be understood by the fact that
the theory is regularized by matrices, and
further examine the dependence of entanglement entropy on the matrix size.
In the interacting case,  
by performing Monte Carlo simulations,
we observe a transition from a generalized volume law, which is obtained by
integrating the square of area law, to the square of area law.
\end{abstract}
\vfill
\end{titlepage}
\vfil\eject

\setcounter{footnote}{0}

\section{Introduction}
\setcounter{equation}{0}
\renewcommand{\thefootnote}{\arabic{footnote}} 
It is widely recognized that noncommutative field theories are deeply connected
to quantum gravity and string theory. On the other hand, 
since the discovery of the Ryu-Takayanagi formula \cite{Ryu:2006bv},
the connection between geometry and quantum entanglement has been revealed.
One can, therefore, expect to gain insight into quantum gravity by studying  quantum entanglement in noncommutative field theories.

Indeed, by studying  a gravity dual of noncommutative super Yang-Mills theory (NCSYM)
proposed in \cite{Hashimoto:1999ut,Maldacena:1999mh}, 
it was conjectured in \cite{Fischler:2013gsa,Karczmarek:2013xxa} that 
entanglement entropy (EE) in NCSYM 
is proportional  to the volume of a focused region
when the volume is small and to the area of the boundary of the region when
the volume is large.
While EE is proportional to the area in ordinary local field theories,
the above volume law in NCSYM would originate 
from the UV/IR mixing \cite{Minwalla:1999px} due to
nonlocal interactions.
Indeed,  in \cite{Shiba:2013jja}, the volume law for EE is obtained in nonlocal theories.

In this paper, we use the words ``volume'' and ``area'' for  real area and length,
respectively, on a sphere.
In \cite{Karczmarek:2013jca,Sabella-Garnier:2014fda}\footnote{For earlier studies, see 
\cite{Dou:2006ni,Dou:2009cw}.}, 
EE in a scalar field theory
on the fuzzy sphere
was studied at zero temperature 
in the free case\footnote{Throughout this paper, the case where
the action consists of only quadratic terms is called  the ``free case'', while
the case where the action includes higher order terms is called the 
``interacting case''.}.
In \cite{OST}, Okuno and the present authors reported the results for EE in
the above theory.
We verified that EE on the fuzzy  sphere in
the free case is proportional
to the square of the area of the boundary, which was suggested 
in \cite{Sabella-Garnier:2014fda}. Moreover, 
we showed the first Monte Carlo results\footnote{For Monte Carlo
studies concerning the fuzzy sphere, see \cite{Azuma:2004zq,Azuma:2005bj,
Medina:2005su,GarciaFlores:2009hf,Panero:2006bx,Das:2007gm}. 
} for the interacting case, where
we found that the behavior of
EE is quite different from that in the free case.
We also found that the finite temperature effect is governed by
the volume law in the interacting case as well as in the free case.
In calculating EE,  
we used a method that is different from the one
in  \cite{Karczmarek:2013jca,Sabella-Garnier:2014fda}.
This method was developed in \cite{Buividovich:2008kq} and used
in \cite{Buividovich:2008kq,Nakagawa:2011su,Itou:2015cyu}.

In this paper, we continue the study of EE in the scalar field theory
on the fuzzy sphere. We present further results for the free case as well as for
the interacting case. In the free case,
we discuss  why EE obeys the square of area law, and
further examine the dependence of EE on the matrix size.
By performing Monte Carlo simulations in the interacting case, we 
observe a transition from a generalized volume law, which corresponds to the integral
of the square of the area, to the square of area law, when
the volume of a focused region is increased.
This phenomenon would be attributed to
the UV/IR anomaly discovered in \cite{Chu:2001xi,CastroVillarreal:2004vh}, which
is the counterpart of the UV/IR mixing in 
field theories on compact noncommutative manifolds.

Another aim of our work is to elucidate geometry in matrix models.
This is in particular important in the context of the study of
matrix models
proposed as nonperturbative formulation of string theory \cite{BFSS,IKKT,DVV}.
Indeed, the above scalar field theory on the fuzzy sphere is realized  in a matrix
model as a regularized theory.
Following a prescription given in
\cite{Karczmarek:2013jca,Sabella-Garnier:2014fda},  we divide
the matrices into two parts, each one corresponding to one of the 
two regions on the sphere.
We would like to see how well this division in the matrix works.


This paper is organized as follows.
In section 2, we review a matrix model that realizes
a scalar field theory on $S^1\times$ fuzzy sphere.
In section 3, we describe how we calculate EE in this theory.
We present the results for EE in the free case in section 4 and in the interacting case
in section 5, respectively.
Section 6 is devoted to conclusion and discussion.

\section{Scalar field theory on the fuzzy sphere}
\setcounter{equation}{0}
\subsection{Scalar field theory on the fuzzy sphere realized by a matrix model}
The commutative counterpart of the noncommutative scalar field theory we consider
in this paper
is defined on $S^1\times S^2$ as
\begin{align}
S_C=\frac{R^2}{4\pi}\int_0^{\beta} dt \int d\Omega \left
(\frac{1}{2}\dot{\phi} (t,\Omega)^2-\frac{1}{2R^2}({\cal L}_i \phi (t,\Omega))^2
+\frac{\mu^2}{2}\phi (t,\Omega)^2
+\frac{\lambda}{4}\phi (t,\Omega)^4 \right) \ ,
\label{commutative action}
\end{align}
where $R$ is the radius of $S^2$, $d\Omega=\sin\theta d\theta d\varphi$ 
is the invariant measure 
for unit sphere,
$\beta$ is the circumference of $S^1$ that corresponds to inverse temperature, and
the dot stands for the derivative with respect to $t$ that parametrizes $S^1$.
${\cal L}_i$ (i=1,2,3) are the orbital angular momentum operators given by
\begin{align}
{\cal L}_{\pm} &\equiv {\cal L}_1 \pm i {\cal L}_2 
=e^{\pm i\varphi} \left( \pm \frac{\partial}{\partial \theta} 
+i \cot \theta \frac{\partial}{\partial \varphi} \right) \ ,\nonumber\\
{\cal L}_3 &= -i \frac{\partial}{\partial \varphi} \ .
\end{align}

To obtain the noncommutative field theory, we replace $S^2$ with the fuzzy sphere
in (\ref{commutative action}).
The resultant theory is realized by a matrix model, that is defined by
\begin{align}
S_{NC}=\frac{R^2}{2j+1}\int_0^{\beta} dt \  \mbox{tr} \left (\frac{1}{2}\dot{\Phi}(t)^2
-\frac{1}{2R^2}[L_i,\Phi (t)]^2 +\frac{\mu^2}{2}\Phi (t)^2+\frac{\lambda}{4}\Phi (t)^4 \right) \ ,
\label{noncommutative action}
\end{align}
where $j$ is a nonnegative integer or a positive half-integer specifying the spin of the representation of $SU(2)$, and  
$\Phi$ is a $(2j+1)\times (2j+1)$ Hermitian
matrix that depends on $t$. $L_i$ are the generators of the SU(2) algebra
for the spin $j$ representation, and
obey the relation
$ [L_i, L_j]=i\epsilon_{ijk} L_k $.

Throughout this paper, we are concerned with the so-called commutative limit
where $j \rightarrow \infty$ with $R$ fixed. As stated below, 
the theory (\ref{noncommutative
action}) agrees with the theory (\ref{commutative action}) at tree level 
in this limit, but differs from it with radiative corrections.
The UV cutoff is given by $\Lambda=\frac{2j}{R}$.
To see the correspondence between the two theories,
it is convenient to introduce
the Bloch coherent 
states $|\Omega \rangle$ $(\Omega=(\theta,\varphi)$) \cite{Gazeau}\footnote{See also
\cite{Alexanian:2000uz,Hammou:2001cc,Presnajder:1999ky,Ishiki:2015saa}.}, which
are localized around a point $(\theta,\varphi)$ on the unit sphere.
The properties of the Bloch coherent state are summarized in appendix A in \cite{OST}.

The explicit form of  $|\Omega \rangle$ is given by
\begin{align}
|\Omega\rangle=\sum_{m=-j}^{j}
\left( 
\begin{array}{c}
2j \\
j+m
\end{array}
\right)^{\frac{1}{2}}
\left( \cos \frac{\theta}{2} \right)^{j+m} \left( \sin \frac{\theta}{2} \right)^{j-m} 
e^{i(j-m) \varphi} |jm\rangle \ ,
\label{explicit form}
\end{align}
where 
$L_{\pm}=L_1 \pm i L_2$ and
$|jm\rangle$ $(m=-j, -j+1,\dots,j)$ are the standard basis 
for the spin $j$ representation of the SU(2) algebra satisfying
\begin{align}
L_{\pm}|jm\rangle &=\sqrt{(j\mp m)(j \pm m+1)}|j m\pm 1\rangle, \nonumber\\
L_3|jm\rangle &= m |jm\rangle \ .
\label{standard basis}
\end{align}
By using (\ref{explicit form}), it is easy to show that
\begin{align}
\frac{2j+1}{4\pi} \int d\Omega \ |\Omega\rangle\langle \Omega | =1
\label{property4}
\end{align}
and 
\begin{align}
|\langle \Omega_1 | \Omega_2 \rangle |^2 = \left( \cos \frac{\chi}{2} \right)^{2j} 
\;\; \mbox{with} \;\; \chi=\arccos (\vec{n}_1\cdot \vec{n}_2) \ ,
\label{inner product}
\end{align}
where $\vec{n}_1=(\sin\theta_1\cos\varphi_1,\sin\theta_1\sin\varphi_1,\cos\theta_1)$
and $\vec{n}_2=(\sin\theta_2\cos\varphi_2,\sin\theta_2\sin\varphi_2,\cos\theta_2)$.
(\ref{inner product}) implies that the width of the Bloch coherent state is given
by $\Delta=\frac{R}{\sqrt{j}}$.

We also introduce the Berezin symbol \cite{Berezin:1974du} defined by
$f_{\Phi(t)}(\Omega)=\langle\Omega|\Phi(t)|\Omega\rangle$,
which is identified 
with $\phi(t,\Omega)$ in (\ref{commutative action}) in the $j\rightarrow\infty$ limit.
First, by using
 (\ref{explicit form}), 
it can easily be shown that 
\begin{align}
f_{[L_i,\Phi] }(\Omega)={\cal L}_i f_{\Phi}(\Omega) \ .
\end{align}
Second, the star product for the two Berezin symbols is defined by
\begin{align}
f_{A}(\Omega)\ast f_{B}(\Omega) \equiv f_{AB}(\Omega)
=\frac{2j+1}{4\pi}\int d\Omega'  \ \langle\Omega | A | \Omega'\rangle
\langle\Omega'|B|\Omega\rangle \ ,
\end{align}
where (\ref{property4}) is used. The star product coincides with the ordinary
one at the tree level in the $j\rightarrow\infty$ limit, while it gives rise to the UV/IR anomaly at the quantum level. 
Indeed, the noncommutative parameter 
is given by $\Theta=\frac{R^2}{4j}$ \cite{Karczmarek:2013jca,Presnajder:1999ky},
which vanishes in the $j\rightarrow \infty$ limit.
Third, by using (\ref{property4}), the trace over 
a matrix is translated into the integral over $S^2$.
In this manner, 
the theory (\ref{noncommutative action}) coincides with
the theory (\ref{commutative action}) at the tree level in the $j\rightarrow \infty$ limit,
while
it differs from the theory  (\ref{commutative action}) 
at the quantum level even in the $j\rightarrow \infty$ limit due to the UV/IR anomaly.
The length scale of nonlocality of the interaction that gives rise to the UV/IR anomaly is given by $\Theta \Lambda \sim R$.

\subsection{Division of the fuzzy sphere}

Following  the prescription in 
\cite{Karczmarek:2013jca}, we divide the fuzzy sphere into two regions.

First, let us see the relationship between the Berezin symbol and 
the matrix elements $\langle jm | \Phi |jm'\rangle$.
We have a relation
\begin{align}
f_{\Phi}(\Omega)=\sum_{m,m'} \langle\Omega | jm\rangle \langle jm' | \Omega\rangle
\langle jm | \Phi | jm'\rangle  \ .
\end{align}
By using (\ref{explicit form}), we find that
\begin{align}
\langle\Omega | jm\rangle \langle jm' | \Omega\rangle
\sim \left(\cos\frac{\theta}{2}\right)^{2j+m+m'}\left(\sin\frac{\theta}{2}\right)^{2j-m-m'}
e^{i(m-m')\varphi}  \ .
\label{Berezin and matrix}
\end{align}
It is easy to show that (\ref{Berezin and matrix}) has 
a sharp peak at \cite{Karczmarek:2013jca}
\begin{align}
\cos\theta = \frac{m+m'}{2j} 
\label{region on sphere and matrix}
\end{align}
with width $\Delta \theta \sim \frac{1}{\sqrt{j}}$.
This observation shows that the matrix elements  $\langle jm | \Phi | j \ n-m\rangle$ 
correspond to the field $\phi$ at $\cos\theta=\frac{n}{2j}$ \cite{Karczmarek:2013jca}.

\begin{figure}[tbp]
\begin{center}
{\includegraphics[width=100mm]{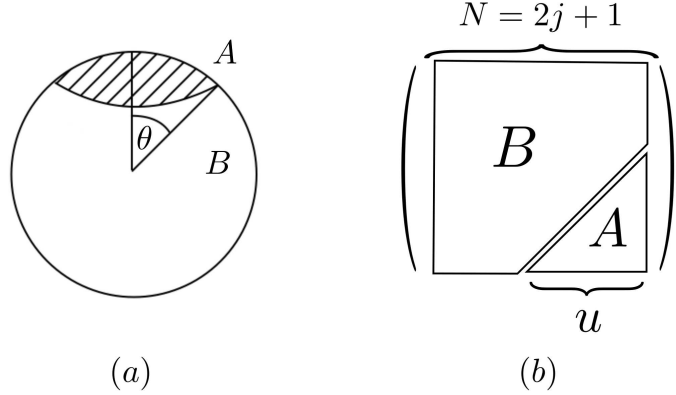}}
\end{center}
\caption{Correspondence between two regions on the fuzzy sphere and in the matrix model.}
\label{sphere and matrix}
\end{figure}

Next, using the relation (\ref{region on sphere and matrix}), we assign  regions A and B
on the sphere in Fig.\ref{sphere and matrix}(a) to parts A and B, respectively,
of the matrix $\Phi$
in Fig.\ref{sphere and matrix}(b).
In order to parametrize region A on the sphere, 
we introduce a parameter $x$, which is the ``volume'' 
of the region A divided by $2\pi R^2$:
\beq
x=1-\cos \theta \ .
\label{x and theta}
\eeq
Note that the ``area'' of the boundary between the regions A and B is given by
\begin{align}
2\pi R \sin\theta = 2\pi R \sqrt{2x-x^2} \ .
\end{align} 
The condition that the element $\langle jm | \Phi | j m' \rangle$
is located 
in part A is given by
\begin{align}
m+m' > 2j-u \ ,
\label{m+m'}
\end{align}
where $u=0,1,2,\dots,4j$. From  (\ref{region on sphere and matrix}), (\ref{x and theta}) 
and (\ref{m+m'}), we find that
\begin{align}
x=\frac{u}{2j} \ .
\label{relation between x and u}
\end{align}

We can put $R=1$  without loss of generality.
In the following sections, we further put $\mu=1$ for simplicity,  
and denote the matrix size $2j+1$ by $N$.

\section{Calculation of EE}
\setcounter{equation}{0}
\subsection{Entanglement entropy}
The division of the matrix $\Phi$ in Fig.\ref{sphere and matrix}(b)
corresponds to decomposing 
the Hilbert space in the theory (\ref{noncommutative action})  to a tensor product
\begin{align}
{\cal H}={\cal H}_A\otimes {\cal H}_B \ .
\label{decomposition of Hilbert space}
\end{align}
EE for subsystem $A$ is defined by
\begin{align}
S_{A}(x)=- \mathrm{Tr} (\rho _{A}\log \rho _{A}) \ ,
\label{definition of entanglement entropy}
\end{align}
where $x=\frac{u}{2j}$.
Here $\rho_A$ is defined by
\begin{align}
\rho_A = \mathrm{Tr}_B(\rho_{\mbox{tot}}) \ ,
\end{align}
where $\mbox{Tr}_B$ stands for the partial trace over ${\cal H}_B$,
and $\rho_{\mbox{tot}}$ is the total density matrix.
We regard $S_A$ as EE for region A in Fig.\ref{sphere and matrix}(a).
Note that as a general property of EE the following relation holds at zero temperature:
\begin{align}
S_A=S_B \ ,
\label{rhoA=rhoB}
\end{align}
which implies that
\begin{align}
&S_A(x)=S_A(2-x) \ , \nonumber\\
&\frac{\partial S_A}{\partial x}(x)=-\frac{\partial S_A}{\partial x}(2-x) \ ,
\label{odd function}
\end{align}
reflecting the symmetry of the system.
The second relation in (\ref{odd function}) will be used in checking the calculation and 
deriving the finite temperature effect.

\subsection{Method to calculate EE}
To calculate EE, we use the method developed in \cite{Buividovich:2008kq}.
This method is based on  the replica method,
in which the definition of EE
 (\ref{definition of entanglement entropy}) is rewritten as
\beq
S_{A}=\lim _{\alpha\rightarrow 1}\left[-\frac{\partial}{\partial \alpha}\mathrm{Tr}\rho_{A}^{\alpha}\right]=\lim _{\alpha\rightarrow 1}\left[-\frac{\partial}{\partial \alpha}\log (\mathrm{Tr}\rho_{A}^{\alpha})\right] \ ,
\label{difS}
\eeq
where $\alpha$ is originally the number of replicas and extended to a real number.

\begin{figure}[t]
 \begin{center}
  {\includegraphics[width=50mm]{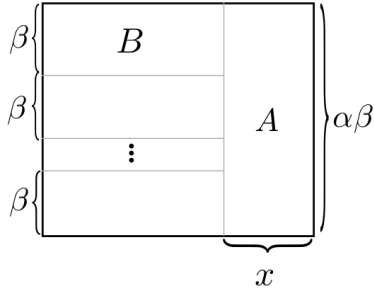}}
 \end{center}
\caption{Replica method.}
\label{fig:replica}
\end{figure}

We introduce $\alpha$ replicas 
$\Phi_n(t)$ $(n=1,\dots,\alpha)$ for $\Phi(t)$ 
in (\ref{noncommutative action}) .
The boundary condition on $\Phi_n(t)$ is depicted in Fig.\ref{fig:replica}:
\begin{align}
\Phi_n(\beta,m,m')&=\Phi_{n+1}(0,m,m')  \;\; \mbox{for part A}  \ , \nonumber\\
\Phi_n(\beta,m,m')&=\Phi_n(0,m,m') \;\; \mbox{for  B} \ ,
\label{boundary condition}
\end{align}
where $n=1,\dots,\alpha$ and $\alpha+1$ is identified with 1 in the first line.
Then, we find a relation
\beq
\mathrm{Tr}\rho^{\alpha}_{A}=\frac{Z(x,\alpha)}{Z^\alpha} \ ,
\label{rho}
\eeq
where $Z(x,\alpha)$ is the partition function of the theory in which
the boundary condition for the replicas is given in (\ref{boundary condition}),
and $Z$ corresponds to the case with $\alpha=1$ and is independent of $x$. 
By substituting (\ref{rho}) into (\ref{difS}),  we obtain
an expression for $S_A$
\beq
S_A(x)=-\lim_{\alpha\rightarrow 1}\frac{\partial}{\partial \alpha} \ln\left(\frac{Z(x,\alpha)}{Z^\alpha}\right) \ .
\eeq
EE for the ground state is given
in the $\beta\rightarrow\infty$ limit, while EE for finite $\beta$
includes the finite temperature effect.

It is much easier to calculate the derivative of $S_A$ with respect to $x$ than
$S_A$ itself: It is expressed as
\beq
\frac{\partial S_A(x)}{\partial x}
=\frac{\partial}{\partial x}\left[-\lim_{\alpha\rightarrow 1}\frac{\partial}{\partial \alpha}\ln\left(\frac{Z(x,\alpha)}{Z^\alpha}\right)\right]=\lim _{\alpha\rightarrow 1}\frac{\partial}{\partial x}\frac{\partial}{\partial \alpha}F(x,\alpha) \ ,
\eeq
where $F(x,\alpha)=-\ln Z(x,\alpha)$. 
Here we approximate the derivative with respect to $\alpha$ as\footnote{
To be conservative, what we calculate is the derivative of 
the R\'enyi entropy with respect to $x$, where the R\'enyi parameter is
equal to 2.}
\begin{align}
&\lim _{\alpha\rightarrow 1}
\frac{\partial}{\partial x}\frac{\partial}{\partial \alpha}F(x,\alpha) \nonumber\\
&\rightarrow 
\frac{\partial}{\partial x}(F(x,\alpha=2)-F(x,\alpha=1)) 
=\lim_{j\rightarrow\infty} \frac{F(x+\varepsilon,\alpha=2)-F(x,\alpha=2)}{\varepsilon}
\ ,
\label{approximation}
\end{align}
where $\varepsilon=\frac{1}{2j}$.


We discretize the time direction with the lattice spacing $a$.

In the free case  where $\lambda=0$, we calculate $F(x,\alpha=2)$
by numerically evaluating the determinant. The method is explained in appendix B
in \cite{OST}.

In the interacting case where $\lambda\neq 0$, we perform Monte Carlo simulation.
We consider an interpolating action
$S_{\mbox{int}}=(1-\gamma)S_{x+\varepsilon}+\gamma S_{x}$, where
$S_{x+\varepsilon}$ and $S_{x}$ are the actions that would give
$F(x+\varepsilon,\alpha=2)$ and $F(x,\alpha=2)$, respectively.
Then the last expression in 
(\ref{approximation}) 
reduces to
\beq
\frac{F(x+\varepsilon,\alpha=2)-F(x,\alpha=2)}{\varepsilon}
=2j\int_0 ^1 d\gamma  \ \langle S_{x+\varepsilon}-S_{x}\rangle _\gamma \ ,
\label{difF}
\eeq
where $\langle \cdots \rangle_\gamma$ represents the expectation value
with respect to the canonical weight $e^{-S_{\mbox{int}}}$.
We take $\gamma$ from 0 to 1 by the step 0.1, 
and calculate $\langle S_{x+\varepsilon}-S_{x}\rangle _\gamma$ for each $\gamma$
by Monte Carlo simulation.
Then,
we finally apply the Simpson formula to the integral over $\gamma$ in 
(\ref{difF}).

\section{Results for the free case}
\setcounter{equation}{0}
\subsection{Behavior of EE in the free case}
\begin{figure}[t]
\begin{center}
 \begin{center}
  {\includegraphics[width=110mm]{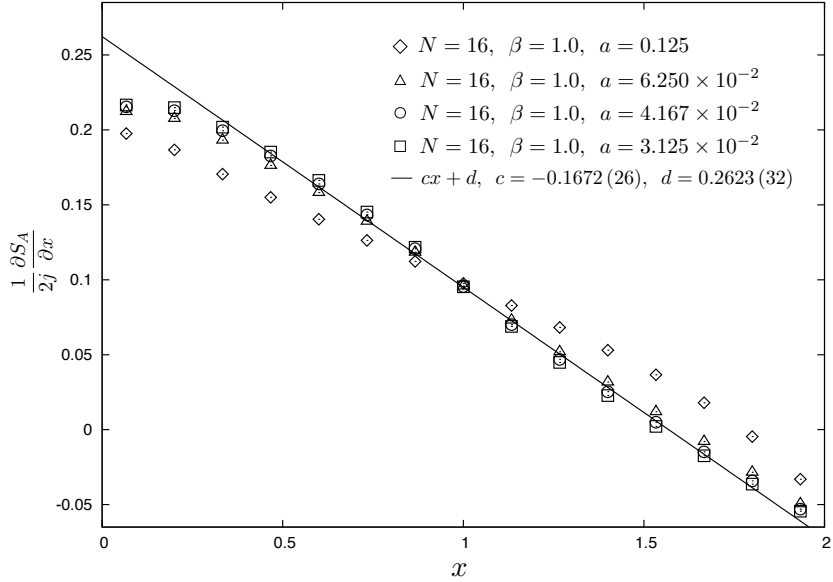}}
 \end{center}
\end{center}
\caption{The quantity $\frac{1}{2j}\frac{\partial S_A}{\partial x}$ 
 at $\lambda=0$, $N=16$ and $\beta=1.0$ is plotted against $x$.
The data for $a=0.125, 6.250\times 10^{-2}, 4.167\times 10^{-2}, 3.125\times 10^{-2}$
are represented by diamonds, triangles, circles, and squares, respectively.
The data for $a=3.125\times 10^{-2}$ are fitted to
$\frac{1}{2j}\frac{\partial S_A}{\partial x}=cx+d$ for
$0.333\leq x \leq 1.800$, which gives 
$c=-0.1672(26)$ and $d=0.2623(32)$. }
\label{beta=1}
\end{figure}

In this subsection, we show our results for the free case ($\lambda=0$)\footnote{
Fig.\ref{beta=1} and Fig.\ref{beta=3,4} were also presented in \cite{OST}.}.
We first calculate $F(x,\alpha=2)$ numerically using the method given in appendix B
in \cite{OST}.
Then, following (\ref{approximation}), we calculate $\frac{\partial S_A}{\partial x}$.



\begin{figure}[t]
\begin{center}
 \begin{center}
  {\includegraphics[width=110mm]{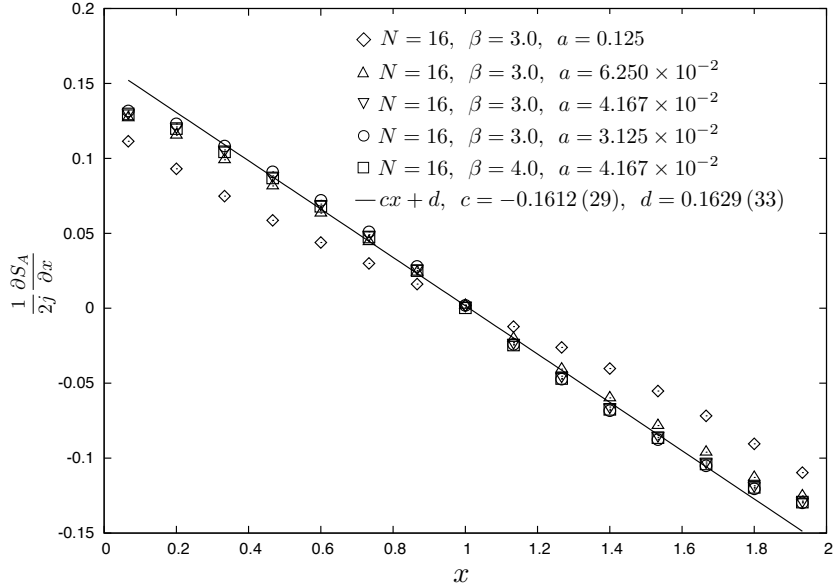}}
 \end{center}
\end{center}
\caption{The quantity $\frac{1}{2j}\frac{\partial S_A}{\partial x}$
is plotted against $x$ at $\lambda=0$ and $N=16$.
The data for $\beta=3.0$ and
$a=0.125, 6.250\times 10^{-2}, 4.167\times 10^{-2}, 3.125\times 10^{-2}$
are represented by 
diamonds, triangles, inverted triangles, and circles, respectively,
while the data for $\beta=4.0$ and $a=4.167\times 10^{-2}$ are represented by
squares.
The data for $\beta=3.0$ and $a=3.125\times 10^{-2}$ are fitted to
$\frac{1}{2j}\frac{\partial S_A}{\partial x}=cx+d$ for
$0.200\leq x \leq 1.800$, which gives 
$c=-0.1612(29)$ and $d=0.1629(33)$. }
\label{beta=3,4}
\end{figure}

\begin{figure}[t]
\begin{center}
 \begin{center}
  {\includegraphics[width=110mm]{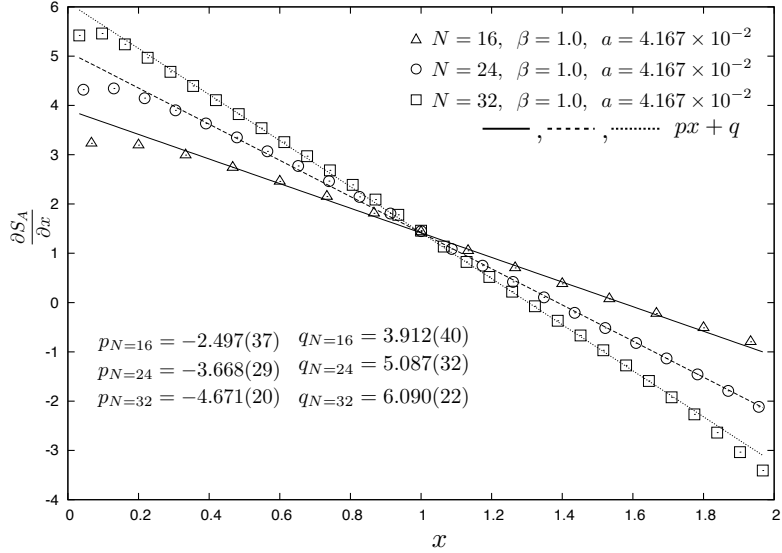}}
 \end{center}
\end{center}
\caption{The quantity $\frac{\partial S_A}{\partial x}$ (not divided by
$2j$) is plotted against $x$ at $\lambda=0$, $\beta=1.0$ and
$a=4.167\times 10^{-2}$.
The triangles, circles, and squares
represent the data for $N=16$, $24$, and $32$, respectively.
The solid line is a fit of the data for $N=16$ to
$\frac{\partial S_A}{\partial x}=px+q$ for $0.333 \leq x \leq 1.667$, which gives
$p=-2.497(37)$ and $q=3.912(40)$. 
The dashed line is a fit of the data for $N=24$ to
$\frac{\partial S_A}{\partial x}=px+q$ for $0.217 \leq x \leq 1.783$, which gives
$p=-3.668(29)$ and $q=5.087(32)$. 
The dotted line is a fit of the data for $N=32$ to
$\frac{\partial S_A}{\partial x}=px+q$ for $0.161 \leq x \leq 1.839$, which gives
$p=-4.671(20)$ and $q=6.090(22)$. 
}
\label{manyN-line}
\end{figure}

\begin{figure}[t]
\begin{center}
 \begin{center}
  {\includegraphics[width=110mm]{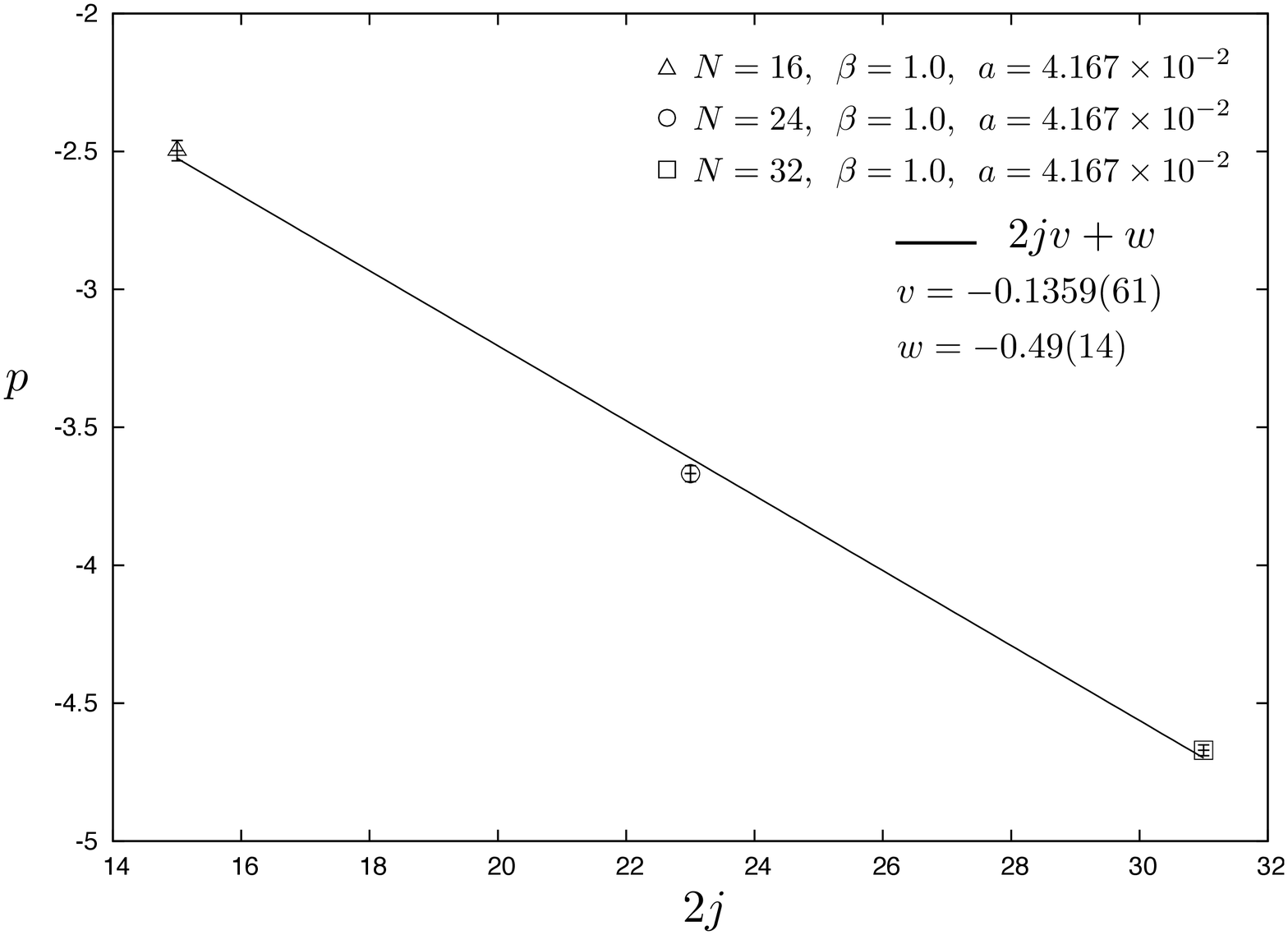}}
 \end{center}
\end{center}
\caption{The values of $p$ obtained in Fig. \ref{manyN-line} are plotted against $2j$.
The solid line is a fit of the data to $p=2jv+w$, which gives $v=-0.1359(61)$ and 
$w=-0.49(14)$.}
\label{manyN-line2}
\end{figure}

We observe that at $N=16$ and $\beta=1.0$ the data for odd $u$ behave smoothly
while the data for even $u$ behave smoothly in a different way
(note that $x=\frac{u}{2j}$).  This difference almost 
disappears at $\beta=4.0$. This difference is considered to originate
from a finite $N$ effect that becomes stronger at high temperature. 
Indeed, we find that the continuum limit in the
time direction can be taken at $N=16$ and $\beta=1.0$ using only the data for odd $u$ or 
even $u$
in such a way that the two continuum limits for odd $u$ and  for 
even $u$ differ only by the finite temperature effect.
We plot only the data
for odd $u$ in what follows..


In Fig. \ref{beta=1}, we plot $\frac{1}{2j}\frac{\partial S_A}{\partial x}$ against $x$.
We plot the data for four values of the lattice spacing $a$  to study
the continuum limit in the time direction at $N=16$ and $\beta=1.0$.
We see that the data for $a=4.167\times 10^{-2}$ and $a=3.125\times 10^{-2}$
almost agree. This implies that $a=4.167\times 10^{-2}$ is close enough to
the continuum limit.
We fit the data for $a=3.125\times 10^{-2}$ to the linear function
$\frac{1}{2j}\frac{\partial S_A}{\partial x}=cx+d$.
In this fitting, we exclude some data points around $x=0$ and $x=2.0$, where 
the volume of 
region A or region B on the sphere is so small that there should be
an ambiguity of the boundary between the two regions due to a finite $N$ effect.
We use the range $0.333\leq x \leq 1.8$ and obtain $c=-0.1672(26)$ and $d=0.2623(32)$.

In Fig. \ref{beta=3,4}, we perform the same analysis at $N=16$ and $\beta=3.0$ as
$N=16$ and $\beta=1.0$. We see that $a=4.167\times 10^{-2}$ is close enough to
the continuum limit also in this case.
Using 
the range $0.2 \leq x \leq 1.8$, we fit the data for $a=3.125\times 10^{-2}$ to
the linear function and obtain $c=-0.1612(29)$ and $d=0.1629(33)$.
Namely, the function is proportional to $1-x$ within the 
fitting error. This function is consistent with (\ref{odd function}). 
This implies that $\beta=3.0$ is close enough to the zero temperature limit
(the $\beta\rightarrow\infty$ limit). 
Indeed,
we also plot the data for $N=16$, $\beta=4.0$ and 
$a=4.167\times 10^{-2}$ in Fig. \ref{beta=3,4}.
The data almost agree with those 
for $N=16$, $\beta=3.0$ and $a=4.167 \times 10^{-2}$. 
This supports the statement that $\beta=3.0$ is close enough to the zero temperature limit.
Thus we find a square of area law at zero temperature:
\begin{align}
S_A \propto 2x-x^2= \sin^2\theta \ .
\label{sin^2theta}
\end{align}

We see that
the difference between 
the two functions $\frac{1}{2j}\frac{\partial S_A}{\partial x}=cx+d$ 
fitted to the data for $\beta=1.0$ and $\beta=3.0$
is almost constant.  This means that the finite temperature effect 
in $S_A$ is proportional to $x$, namely the volume 
of region A. This volume law for the finite temperature effect
is in general seen in local field theories.
Fitting the data with even $u$ for $N=16$, $\beta=1.0$ and
$a=3.125 \times 10^{-2}$ to $\frac{1}{2j}\frac{\partial S_A}{\partial x}=cx+d$
for $0.133 \leq x \leq 1.6$ gives
$c=0.1626(26)$ and $d=0.2690(22)$. As we stated, 
the differences between the data for odd $u$ and the data for even $u$
arise in the finite temperature effect.

In Fig. \ref{manyN-line}, we plot $\frac{\partial S_A}{\partial x}$ (not divided by $2j$)
at $\beta=1.0$, $a=4.167\times 10^{-2}$, and $N=16, 24, 32$.
against $x$ to examine the large-$N$ (large-$j$) limit, which corresponds to
the continuum limit of the fuzzy sphere.
We see that the data for all $N$'s coincide at $x=1$
and that each data can be fitted 
to the linear function $\frac{\partial S_A}{\partial x}=px+q=-p(1-x)+p+q$, where $p+q$ 
is independent of $N$.
This is consistent with the above observation that the finite temperature effect 
is proportional to the volume, and further implies that the finite temperature effect is
independent of $N$. We exclude a shorter range of $x$ in fitting the data as $N$ increases.
This supports the statement that the ambiguity of the boundary is a finite $N$ effect
so that it vanishes in the $N\rightarrow\infty$ limit.
As we stated,
we observe that the difference between the data for odd $u$ and the data for even $u$
becomes smaller as $N$ increases.

In Fig. \ref{manyN-line2}, we plot the values of $p$ obtained in 
Fig.\ref{manyN-line} against $2j$. We obtain a good fit of  the data to $p=2jv+w$.

To summarize, we find that in the free case EE behaves
in the $a\rightarrow 0$ limit with large $j$ as
\begin{align}
S_A =\left(|v|j + \frac{|w|}{2}\right) \sin^2\theta + g (1-\cos\theta) \ ,
\label{behavior of EE in free case}
\end{align}
where $v$ and $w$ are constants independent of $j$ and $\beta$, and 
$g$ is a constant\footnote{$g=p+q$.} independent of $j$ 
and vanishes in the $\beta\rightarrow\infty$
limit. Namely, the first term corresponds to EE at zero temperature, which
is proportional to the square of area and depends on the UV cutoff $j$.
The second term corresponds to the finite temperature effect.
It is governed by the volume law and independent of the UV cutoff $j$.
The $j$ dependence of the first term in (\ref{behavior of EE in free case}) 
is observed in \cite{Karczmarek:2013jca,Sabella-Garnier:2014fda}
and the $\theta$ dependence in the first term is suggested in \cite{Sabella-Garnier:2014fda}.
Thus we verified the behavior of EE at zero temperature by using 
a method different from the one in  \cite{Karczmarek:2013jca,Sabella-Garnier:2014fda}.
This indicates the validity of our method to calculate EE.


\subsection{Origin of the square of area law}
In local field theories, the leading contribution to EE of a focused region A
at zero temperature obeys 
the area law. Namely, it is
proportional to $|\partial A|/\epsilon^{d-1}$, where $|\partial A|$ is
the area of the boundary of the region A, $\epsilon$ is a UV cutoff, and $d$ is the space
dimension.
This behavior is understood from the fact that region A interacts with
the outside through the boundary in local field theories.
On the other hand,
we have confirmed that the leading contribution to EE at zero temperature in the free case
 is proportional to
the square of area, namely $N\sin^2\theta$, although we had naively expected it
to obey the area law.
Because this depends on the UV cutoff $N$, we need to go back to a regularized 
theory (\ref{noncommutative action}) to discuss the origin of this square of area law.

For $\lambda=0$, the matrix model action (\ref{noncommutative action}) is local
with respect to the matrix elements because $L_i$ are tridiagonal in the standard
basis (\ref{standard basis}).
Thus the degree of freedom in the boundary between
parts A and B 
of the matrix $\Phi$
in Fig.\ref{sphere and matrix}(b) should contribute to EE at zero temperature.
The number of states $|jm\rangle$ that effectively contribute to the degree of
freedom in the boundary would
be proportional to the area of the boundary divided by the width of the Bloch
coherent states $\Delta=\frac{R}{\sqrt{j}}$:
\begin{align}
2\pi R\sin\theta \times \frac{1}{\Delta} = 2 \pi  \sqrt{j} \sin\theta \ .
\end{align}
Moreover, the matrix elements 
$\langle jm |\Phi | j m'\rangle$ are bilocal in the sense that they have two
indices $m$ and $m'$.  It is therefore natural that the leading contribution to EE at zero temperature
is proportional to 
\begin{align}
(\sqrt{j} \sin\theta) ^2 \sim N \sin^2\theta \ .
\end{align}

\section{Results for the interacting case}
\setcounter{equation}{0}


\begin{figure}[t]
\begin{center}
 \begin{center}
  {\includegraphics[width=110mm]{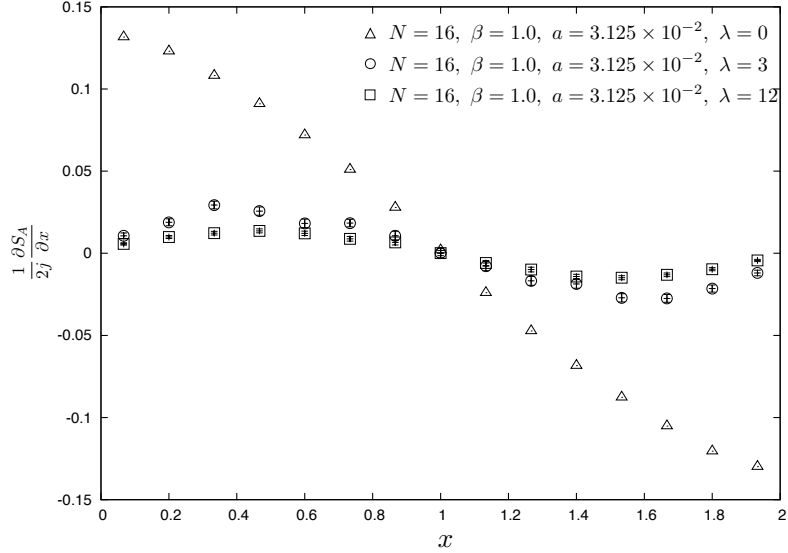}}
 \end{center}
\end{center}
\caption{After the value at $x=1$
is subtracted from the data, 
$\frac{1}{2j}\frac{\partial S_A}{\partial x}$
is plotted against $x$ at $N=16$, $\beta=1.0$, and $a=3.125\times 10^{-2}$.
The triangles, circles, and squares
represent the data for $\lambda=0$, $3.0$, and $12.0$, respectively.
}
\label{three lambda}
\end{figure}

\begin{figure}[t]
\begin{center}
 \begin{center}
  {\includegraphics[width=110mm]{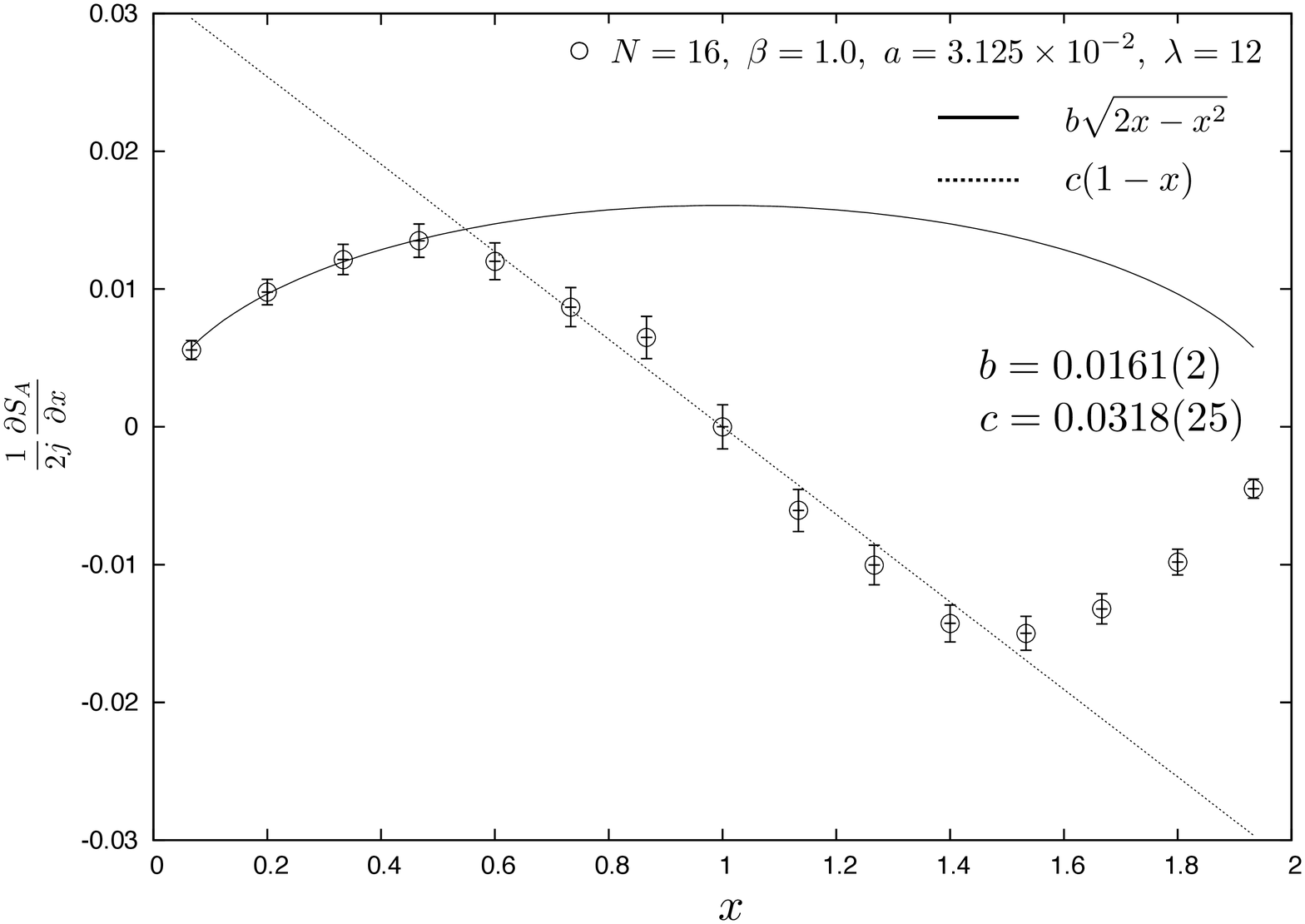}}
 \end{center}
\end{center}
\caption{After the value at $x=1$ is subtracted from the data,
$\frac{1}{2j}\frac{\partial S_A}{\partial x}$ is plotted against $x$ 
at $\lambda=12.0$, $N=16$, $\beta=1.0$ and $a=3.125\times 10^{-2}$.
The solid line is a fit of the first four data points 
($x=0.067 \sim 0.467$)  to $\frac{1}{2j}\frac{\partial S_A}{\partial x}
=b\sqrt{2x-x^2}$, which gives $b=0.0161(2)$. 
The dotted line is a fit of the next four data points ($x=0.6 \sim 1.0$) to 
$\frac{1}{2j}\frac{\partial S_A}{\partial x}=c(1-x)$, which gives $c=0.0318(25)$.
}
\label{lambda=12fit}
\end{figure}


In this section, we study the interacting case.
We perform Monte Carlo simulations at $N=16$, $\beta=1.0$ and
$a=3.125\times 10^{-2}$.


We again see a difference between odd $u$ and even $u$, 
similar to the free case. We plot only the data for odd $u$ in the interacting case also.

In Fig. \ref{three lambda}, we plot $\frac{1}{2j}\frac{\partial S_A}{\partial x}$ 
in the interacting case together with that in the free case against $x$.
We plot the data for $\lambda=3.0, 12.0$ (the interacting case) and for
$\lambda=0$ (the free case)
after we subtract each value at $x=1$ from the data for each $\lambda$.
We see that the shifted data are consistent with (\ref{odd function}).
This suggests that
the finite temperature effect  in $S_A$ is also proportional to the volume in
the interacting case.
The shape of the data in the interacting case is different from
that in the free case, which should be attributed to non-locality of the interaction.
Moreover, the magnitude in the interacting case is quite smaller than that in the 
free case. 

As we stated in section 2.1, the length scale of nonlocality of the interaction
is of order $R=1$.
Therefore we can, in general, expect that
there is a transition from the ``volume law'' to the square of area law,
where the ``volume law'' is given by the integral of the square of the area.
We assume that the transition happens around $\theta=\theta_0$.
Namely, for $\theta_0 \lesssim \theta \leq \frac{\pi}{2}$ EE behaves as 
the square of area law
\begin{align}
S_A= j c \sin^2\theta \ ,
\end{align}
which leads to
\begin{align}
\frac{1}{2j}\frac{\partial S_A}{\partial x}= c(1-x) \ ,
\label{square of area law}
\end{align}
while
for $0 \leq \theta \lesssim \theta_0$ EE behaves as
\begin{align}
S_A= j c \int_0^{\theta} \sin^2\theta' d\theta' \ ,
\end{align}
which leads to
\begin{align}
\frac{1}{2j}\frac{\partial S_A}{\partial x}= \frac{c}{2} \sqrt{2x-x^2} \ .
\label{volume law}
\end{align}

In Fig. \ref{lambda=12fit}, we again plot the data for $\lambda=12.0$ and
fit the first four data points ($x=0.067 \sim 0.467$) to $\frac{1}{2j}\frac{\partial S_A}{\partial x}= b \sqrt{2x-x^2}$ and 
the next four data points ($x=0.6 \sim 1.0$) to $\frac{1}{2j}\frac{\partial S_A}{\partial x}= c(1-x)$.
We obtain $b=1.61\times 10^{-2}(2)$ and $c=3.18\times 10^{-2}(25)$.
These values are consistent with (\ref{square of area law}) and (\ref{volume law}).
We apply the same analysis to the data for $\lambda=10.0$ and
again obtain
a good fit consistent with (\ref{square of area law}) and (\ref{volume law}).

\section{Conclusion and discussion}
\setcounter{equation}{0}
In this paper, we calculated EE in the scalar field theory
on the fuzzy sphere using the method developed
in \cite{Buividovich:2008kq}.
In the free case, we confirmed that EE at zero temperature is proportional to
the square of the area of the boundary and the leading contribution to it is proportional
to the UV cutoff $N$.
We discussed the reason for this peculiar law.
These behaviors are consistent with the observations 
in \cite{Karczmarek:2013jca,Sabella-Garnier:2014fda}. 
We also found that the finite temperature effect in EE 
is proportional to the volume and independent
of the UV cutoff $N$. This property of the finite temperature effect
is shared with ordinary local field theories.

In the interacting case, 
we performed  Monte Carlo simulations to calculate EE. 
We found that the magnitude of EE in the interacting case is quite small
compared to that in the free case. We observed  a transition from
the ``volume law'' to the square of area law for EE as the volume is increased.
The former is obtained by integrating the latter over $\theta$.
This transition should originate from nonlocal nature of the interaction.
We also found that the finite temperature effect in EE is proportional to the volume
in the interacting case.

The fact that we can interpret our results geometrically following the division of the 
sphere in Fig.\ref{sphere and matrix}(a) shows that the division of the matrix 
in Fig.\ref{sphere and matrix}(b)
works well.

In the interacting case, we need to study renormalization and the continuum limit.
Indeed, at
$N=16$, $\beta=1.0$, and $a=3.125\times 10^{-2}$,
we obtained a good fit to the transition from the ``volume law'' to
the square of area law for $\lambda=10.0, 12.0$ but not for $\lambda=3.0$.
This suggests that the former is closer to the continuum limit than the latter.

In \cite{Sabella-Garnier:2014fda}, it was shown that 
mutual information (MI) in the free case agrees with that in the ordinary scalar field
theory on $R\times S^2$. This result is reasonable because
MI is independent of the UV cutoff and the matrix model
(\ref{noncommutative action}) with $\lambda=0$ reduces to the ordinary scalar
field theory on $S^1\times S^2$ in the limit where the UV cutoff goes to infinity.
We expect differences in the interacting case  
between MI in the noncommutative theory and that in the ordinary theory.

We hope to report progress in the above issues in the near future.

\section*{Acknowledgements}
We would like to thank Shizuka Okuno for collaboration at an early stage of this work.
Numerical computation was carried out on SR16000 and XC40 at YITP in Kyoto University
and SR16000 and FX10 at the University of Tokyo.
The work of A.T. is supported in part by Grant-in-Aid
for Scientific Research
(nos. 24540264, 23244057, and 15K05046)
from JSPS.


\end{document}